From the editor: Can the right requirements boost developer satisfaction and happiness? We believe they can. In keeping with this issue's theme "Wellbeing for Resilience: Developers Thrive," we discuss the connection between the three keywords wellbeing, resilience, and thriving. How could requirements engineering foster these qualities? While there hasn't been much research on this topic, we see opportunities for future work. Let's initiate the discussion!

# Requirements for Organizational Resilience: Engineering Developer Happiness

**Markus Borg and Daniel Graziotin**

Bob is again leaning backward in his chair, a bead of sweat trickling down his temple as he stares at the mess of code on his three screens. It's on him to adapt this part of the gnarled legacy system for the new feature that has been discussed for weeks. It's Thursday afternoon, time to go home, time to be a parent and not just a developer. Yet here he is, wrestling with a codebase that's become a testament to a decade of shortcuts. The changes should be pushed before the weekend. How did it come to this? Bob looks at the clock again, teeth are clenching, giving that recent recruiter's headhunting email another thought. Would the code be greener on the other side?

## Happiness and Wellbeing

Bob's frustrating experience is not unique, but a reflection of a broader challenge within the software industry. This issue's theme is well-being for resilience in the world of software developers. Developers' well-being is foundational to their productivity and their ability to cope with stress. And when they thrive, they will probably resist the lure of new opportunities! This column explores the links from happiness to wellbeing to resilience – and discusses what role requirements can play in this.

Yet, diving into the philosophical and social science realms reveals a compelling argument for reevaluating what truly contributes to our happiness and, by extension, our effectiveness at work. Haybron presents a thought-provoking perspective on happiness, one that goes beyond the fleeting highs of a job well done or the superficial satisfaction with our career trajectory [Hay05]. He argues that at the heart of genuine happiness is our emotional state - how we feel on a deeper, more sustained level.

For us in the software industry, Haybron's insights serve as a reminder that nurturing an environment that supports emotional well-being could be vital to not just personal satisfaction but also professional excellence. Emotional health isn't just a nice-to-have; it's integral to how happy and productive we are in our roles. So, as we chase after the next milestone or sprint, it's worth considering: are we also paying attention to the emotional climate of our teams? Haybron's research suggests that a happier team, in the truest sense of emotional well-being, could very well be our most valuable asset.

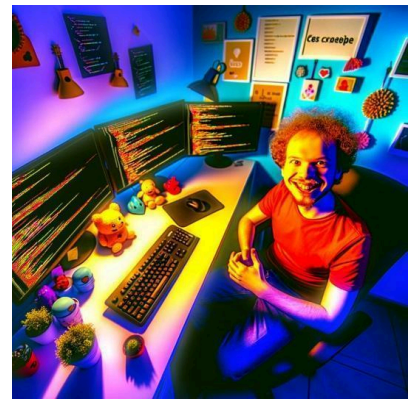

Figure SEQ Figure \\* ARABIC 1: *Bob – when happy.*

## Wellbeing and Resilience

The happiness-wellbeing connection extends to resilience. On a personal level, a happy developer is also better equipped to face the hardships of development. Periods of stress are normal, but we all know that too much of it is detrimental. Already eXtreme Programming stressed programmer welfare with "sustainable pace" as a fundamental practice. This was expressed as developers should not work more than 40-hour weeks, and if there is one week of overtime, then the next week should not include more overtime. Developers perform at their best and most creatively if they are well-rested.

On the organizational level, the connection between wellbeing and resilience is equally important. The tech industry's rapid pace and the global developer shortage mean talent retention is critical. Unhappiness, as shown in Bob's contemplation of a job switch, leads to turnover, knowledge loss, and the high costs of training newcomers. Developer churn can be devastating, especially if key people decide to leave – discussions about the *bus factor* must be taken seriously [Jab22].

In this context, we argue that Requirements Engineering (RE) could take on a new dimension. Traditionally focused on user impact, we're instead interested in how they affect developers' work environments and, by extension, their happiness and productivity. What is your experience? Have you observed situations where the requirements directly impacted the well-being of your development team? We propose addressing requirements with an eye toward developer well-being for a sustainable development practice.

## Requirements and Happiness?

Previous RE research has targeted emotions, but the focus has been on the users. For example, in a rich paper published 20 years ago, Ramos et al. argued that RE needs to consider users' emotional issues to support acceptance of company-wide information systems [Ram05]. They also presented guidelines on how to elicit users' emotions and beliefs.

Another research direction with several related contributions deals with requirements for video games. Unsurprisingly, since video games are all about the gamers' experience and their emotions. A good example of this strand of work is the concept of emotional requirements to express desired emotions in gamers [Cal06].

Substantial research also exists on affective computing, that is, "machines that process, interpret, simulate, and analyze human behavior and emotions." In the RE community, the international workshop series AffectRE has hosted five instances. However, the workshop has focused on the use of affective computing with external users from an RE perspective, rather than how requirements cause affects (emotions and attitudes) in the internal developers.

What is the impact of requirements on the developers' happiness at work? To what extent can requirements be specified to make developers happy? And even if not directly, is there a chain that starts with requirements and ends with shipped software, which, at any moment, influences how developers feel? Is it possible to tailor requirements in a way that not only fulfills project objectives but also nurtures developer happiness? Many questions.

The closest work we have identified is from Colomo-Palacios and colleagues [Col10]. The authors use the *circumplex model of emotion* to investigate how both users and developers emotionally perceive the evolution of a requirements specification. They found that developers and users feel more pleasure but less arousal in the later versions of the specification. This phenomenon suggests that, much like a well-aged relationship, a maturing requirements specification offers a deeper sense of satisfaction, even if the initial excitement fades.

## Quality Requirements and Feeling Proud

Quality in software is a source of pride for developers. Working on a project with high-quality requirements not only enhances the end product but also boosts the morale of those crafting it. Would you be happy if you developed a product or service of subpar quality? No developer would like to put their name on a solution with terrible UX, poor performance or low reliability. What about developing an insecure system with vulnerabilities? It would provoke feelings of professional embarrassment rather than achievement.

It's well-established that quality requirements are crucial for the success of software products in today's competitive market. On top of that, they are also vital for the happiness of the developers themselves. Among the many facets of quality, our previous work shows how important code quality is for the developers. Bad code quality and coding practice are leading causes of developer unhappiness [Gra17]. Moreover, unhappiness leads to low productivity, low code quality, low motivation, and work withdrawal [Gra18]. See the vicious circle here?

No developer enjoys the task of maintaining and evolving code weakened by years of hotfixes. Developers' desire to work on clean, efficient code isn't just a preference— it appears to be a driving force. Faced with frustration and high market demand for their skills, developers like Bob are inclined to switch employers.

## Breaking the Vicious Cycle

We find ourselves facing a vicious cycle that exacerbates the challenges within the tech sector. A low-quality codebase not only diminishes developer satisfaction but also contributes to the industry's high churn rate.

Figure 2 illustrates the cycle. As discussed, poor code (A) makes developers unhappy (B). With supply and demand at play, unhappy developers often seek new horizons (C). To compensate, new developers must be onboarded. Our previous research [Bor23a] has shown that integrating new developers into subpar codebases is slow (D). Additionally, evidence suggests that newly onboarded developers – with lower levels of code *ownership* – are more prone to introducing code issues [Bir11], thus further degrading code quality (E). This creates a downward spiral, leading to an environment where the initial problems only intensify.

So, what's the way out of this detrimental cycle? We posit that RE holds a key role in addressing these challenges. Imagine the potential shift if we were to implement explicit quality requirements for the codebase itself. The center of Figure 2 shows such an intervention. Measuring code quality isn't about just pleasing your developers with vanity metrics – although some might like getting positive feedback on their software craftsmanship – using code quality KPIs can be a business-critical decision to break the vicious cycle. We suggest that you specify your code quality targets, continuously monitor your current level, and move toward your goals [Bor23b]. This will lead to happier developers – and a sustainable development organization with loyal employees.

Somewhere, in an alternate universe where quality reigns supreme, Bob finds himself in a realm of coding excellence. Here, he is not just a developer; he is a custodian of quality, proud and committed, with no reason to look for other opportunities. The grass, err, the code, is greener on this side already! On the organizational level, there will be an increased resilience with less developer churn. The occasional career moves still happen, but Bob and his senior colleagues quickly show the newly onboarded team members the ropes leading to great code. On top of that, there is another effect. Business value. High-quality code can evolve faster to meet new needs on the market. Together with happy, productive developers – this is a killer combo!

Do you have any thoughts about the connection between requirements and happiness? We've yet to discuss functional requirements! Consider the impact of how aligning – or not – with developers' personal values impacts their fulfillment. Imagine an environmentalist coding for green energy versus oil and gas projects. What about addictive video games? Online casinos? Military systems? Passions and morale compasses vary widely. We'd love to hear about your thoughts on such matters!

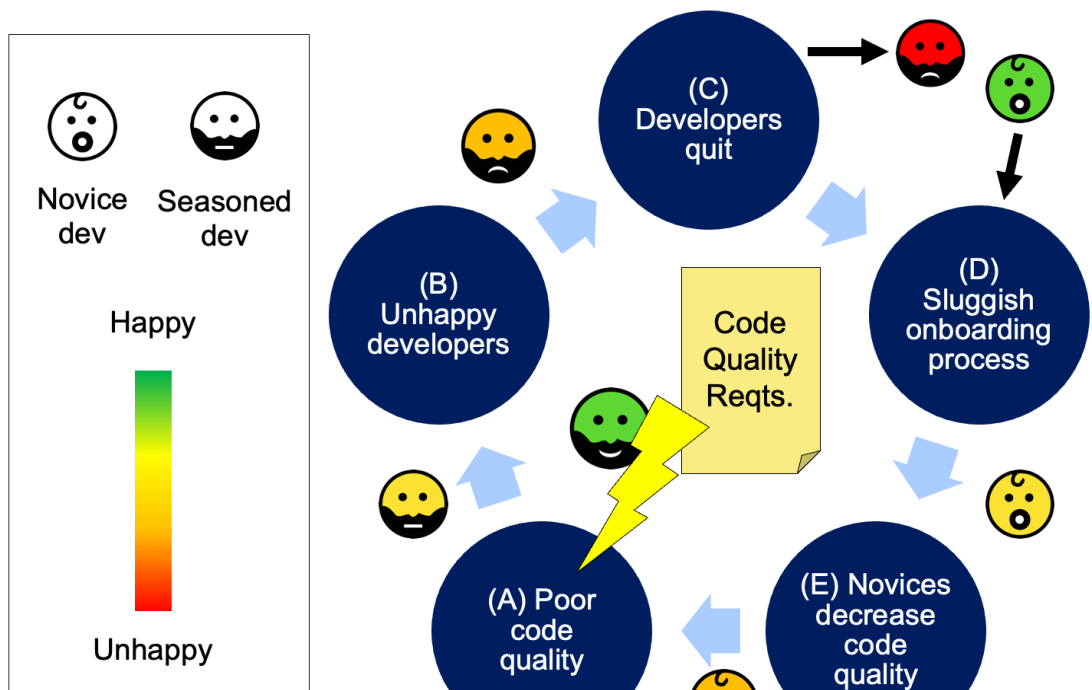

## References

[Hay05] Haybron, D. M., "On Being Happy or Unhappy," *Philosophy and Phenomenological Research*, 71(2), 2005, pp. 287-317.

[Jab22] Jabrayilzade, E., Evtikhiev, M., Tüzün, E. and Kovalenko, V., "Bus factor in practice," In *Proc. of the 44th International Conference on Software Engineering: Software Engineering in Practice*, 2022, pp. 97-106.

[Cal06] Callele, D., Neufeld, E. and Schneider, K., "Emotional requirements in video games," In *Proc. of the 14th IEEE International Requirements Engineering Conference*, 2006, pp. 299-302.

[Col10] Colomo-Palacios, R., Hernández-López, A., García-Crespo, Á. and Soto-Acosta, P. "A study of emotions in requirements engineering," In *Organizational, Business, and Technological Aspects of the Knowledge Society: Third World Summit on the Knowledge Society*, 2010, pp. 1-7. Springer Berlin Heidelberg, 2010.

[Bor23a] Borg, M., Tornhill, A. and Mones, E., "U owns the code that changes and how marginal owners resolve issues slower in low-quality source code," In *Proc. of the 27th International Conference on Evaluation and Assessment in Software Engineering*, 2023, pp. 368-377.

[Bor23b] Borg, M., !Requirements on Technical Debt: Dare to Specify Them!," *IEEE Software*, 40(2), 2023, pp.8-12.

[Ram05] Ramos, I., Berry, D.M. and Carvalho, J.Á., "Requirements engineering for organizational transformation," *Information and Software Technology*, 47(7), 2005, pp. 479-495.

[Bir11] Bird, C., Nagappan, N., Murphy, B., Gall, H. and Devanbu, P. "Don't touch my code! Examining the effects of ownership on software quality," In *Proc. of the 19th ACM SIGSOFT symposium and the 13th European conference on Foundations of software engineering*, 2011, pp. 4-14.

[Gra17] Graziotin, D., Fagerholm, F., Wang, X. and Abrahamsson, P., "On the unhappiness of software developers," In *Proc. of the 21st international conference on evaluation and assessment in software engineering*, 2017, pp. 324-333.

[Gra18] Graziotin, D., Fagerholm, F., Wang, X. and Abrahamsson, P., "What happens when software developers are (un)happy," *Journal of Systems and Software*, 140(6), 2018, pp. 32-47.

## About the authors

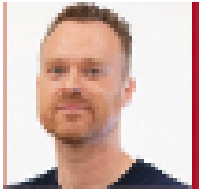

Markus Borg is a principal researcher at CodeScene, 215 32 Malmö, Sweden and an adjunct associate professor at the Department of Computer Science, Lund University, Sweden. Contact him at markus.borg@codescene.com

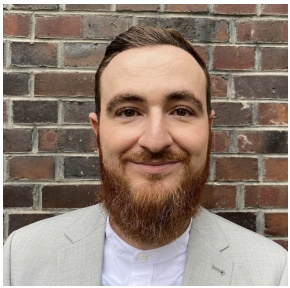

Daniel Graziotin is a full professor at the University of Hohenheim, 70599 Stuttgart, Germany. Contact him at graziotin@uni-hohenheim.de

**Abstract for IEEE Explore. 45 words or less.**

Happy developers are more productive and stay longer. Previous work shows a strong link between poor code quality and unhappiness, which might be self-perpetuating. We argue that code quality requirements could be used to break a vicious cycle – and support organizational resilience.